\documentclass{aastex}          
\usepackage{spr-astr-addons}    
\usepackage{url}

\begin{document}
%
\title{Problems and possibilities in fine-tuning 
of the Cepheid {\boldmath $P$-$L$} relationship}

\shorttitle{The dispersion of the Cepheid $P$-$L$ relationship}
\shortauthors{Szabados \& Klagyivik}

\author{L. Szabados\altaffilmark{}} 
\and 
\author{P. Klagyivik\altaffilmark{}}
\affil{Konkoly Observatory, Budapest, Hungary}
\email{szabados@konkoly.hu} 


\begin{abstract}
Factors contributing to the scatter around the ridge-line 
period-luminosity relationship are listed, followed by a 
discussion how to eliminate the adverse effects of these factors
(mode of pulsation, crossing number, temperature range, reddening,
binarity, metallicity, non-linearity of the relationship, blending),
in order to reduce the dispersion of the $P$-$L$ relationship.
\end{abstract}


\section{Introduction}
\label{s:intro}

\begin{figure}[!hb]
\includegraphics[width=\columnwidth]{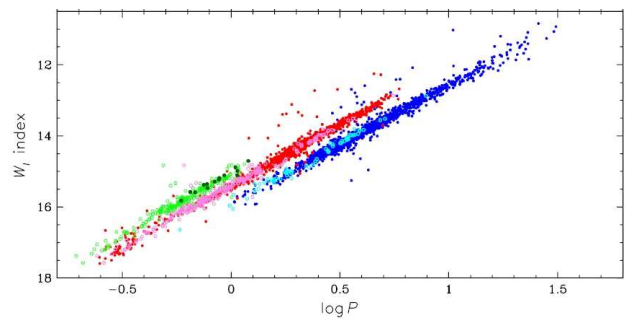}
\caption{Empirical $P$-$L$ relationship for Cepheids in the
Large Magellanic Cloud. To decrease the scatter, Wesenheit magnitudes
are indicated along the vertical axis \citep{Setal08}. Blue and cyan 
points show fundamental-mode pulsators, red and magenta denote
first-overtone, green characters second overtone pulsators. 
Solid dots are single-mode Cepheids, empty circles represent 
the respective mode of double-mode Cepheids.
The dispersion is still about 0\fm5.}
\label{fig:empplr}
\end{figure}

The century-long history of the period-luminosity ($P$-$L$)
relationship of Cepheids can be characterised as a permanent attempt
at improving the calibration of this relationship fundamental for
securing the cosmic distance scale. In spite of tremendous effort, 
the distribution of the points in the $P$-$L$ plots derived for 
any galactic system shows a wider dispersion along the ridge-line
approximation than hoped for (Fig.~\ref{fig:empplr}). The current 
situation is summarized concisely by Marconi (\citeyear{M09}) and, 
in a broader context -- from the point of view of determining the 
Hubble constant --, by \citeauthor{FM10} (\citeyear{FM10}).

Lately, most works on the $P$-$L$ relationship deal with
the metallicity dependence of the luminosity of Cepheids and the
possible non-linearity of the relationship. There are, however, 
other factors contributing to the dispersion of the empirical 
$P$-$L$ relation. 
The aim of this paper is to list all these factors and discuss
the elimination of their adverse effect.

\section{Causes of the dispersion of the $P$-$L$ relationship}
\label{s:causes}

The following factors contribute to the width of the empirical
$P$-$L$ plot:\\
\noindent -- pulsation mode,\\
\noindent -- crossing number (evolutionary stage),\\
\noindent -- effective temperature of the star,\\
\noindent -- interstellar reddening,\\
\noindent -- binarity,\\
\noindent -- blending,\\
\noindent -- metallicity,\\
\noindent -- helium content,\\
\noindent -- nonlinearity of the relationship,\\
\noindent -- other (magnetic field; overshooting; 
depth effect in the host galaxy; etc.).

For individual factors, the amount of the effect can vary 
from negligible to considerable and can depend on the 
pulsation period.
   
\section{Reduction of the dispersion}
\label{s:reduct}

Existence of the $P$-$L$ relationship follows from basic physics
(eigenfrequency of a radial oscillator and Stefan-Boltzmann law
-- see \citeauthor{FM10} (\citeyear{FM10}) and references therein)
which makes Cepheids standard candles for establishing the cosmic
distance scale. The reduction of the scatter in the empirically
determined $P$-$L$ relationship is, therefore, an obvious goal
in the calibration procedure. However, each `widening' effect 
has to be dealt with separately.

\subsection{Pulsation mode}
\label{ss:mode}

For a given star performing radial pulsation, the oscillation
in the fundamental mode has the longest period, while 
oscillations in overtones are characterised by shorter period 
-- without any change in stellar luminosity. 
This behaviour gives rise to a separate sequence
for each pulsation mode which is a conspicuous feature in the
$P$-$L$ plot of extragalactic Cepheids (see Fig.~\ref{fig:empplr}).

In the case of external galaxies, the different sequences show up 
provided the plot is based on a rich Cepheid sample. Assuming
a common distance for all Cepheids in a given galaxy, the
pulsation mode of the individual Cepheids can be inferred from
the period vs. apparent brightness diagram plotted for Cepheids
because this diagram reflects the structure of the $P$-$L$ plot
(see Fig.~\ref{fig:empplr}).

\begin{figure}[!b]
\includegraphics[width=\columnwidth]{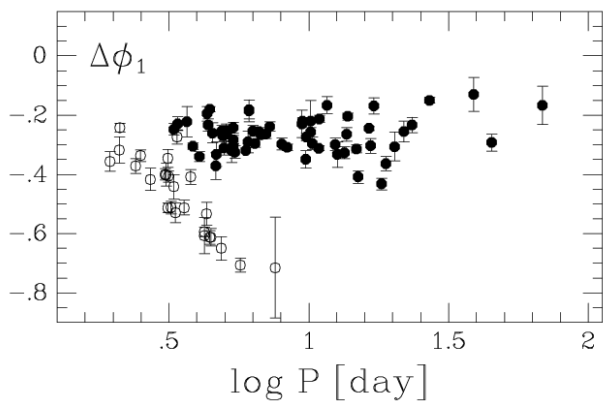}
\caption{The phase lag as defined in the text is a reliable
criterion for the mode identification \citep{Oetal00}}
\label{fig:mode}
\end{figure}

In the case of Galactic Cepheids, however, the period vs. apparent 
brightness diagram is senseless because of the huge range of
distance of Cepheids involved. The pulsation mode has to be
determined before deriving the luminosity. Fourier decomposition
of the photometric and radial velocity phase curves can be
efficiently used for the mode determination: values of the Fourier 
parameters $R_{21}$, $\phi_{21}$, $R_{31}$, and $\phi_{31}$ 
determined from the brightness variations are sufficient to assign 
the pulsation mode of any Cepheid. In those period intervals where
$\phi_{21}$ values overlap for different modes, $\phi_{31}$ values
offer good selection criteria \citep{Setal10}.

If reliable radial velocity phase curve is also available for a
Cepheid, then the period dependent difference of the Fourier phases
 $\Delta \phi_1 = \phi_{21}^{\rm v_{rad}} - \phi_{21}^{\rm phot}$
serves as an indicator of the pulsation mode \citep{Oetal00}.
Usefulness of this empirical phase lag parameter for mode identification
(see Fig.~\ref{fig:mode}) was confirmed theoretically by \citet{Szetal07}.

\subsection{Crossing number}
\label{ss:crn}

\begin{figure}[!b]
\includegraphics[width=\columnwidth]{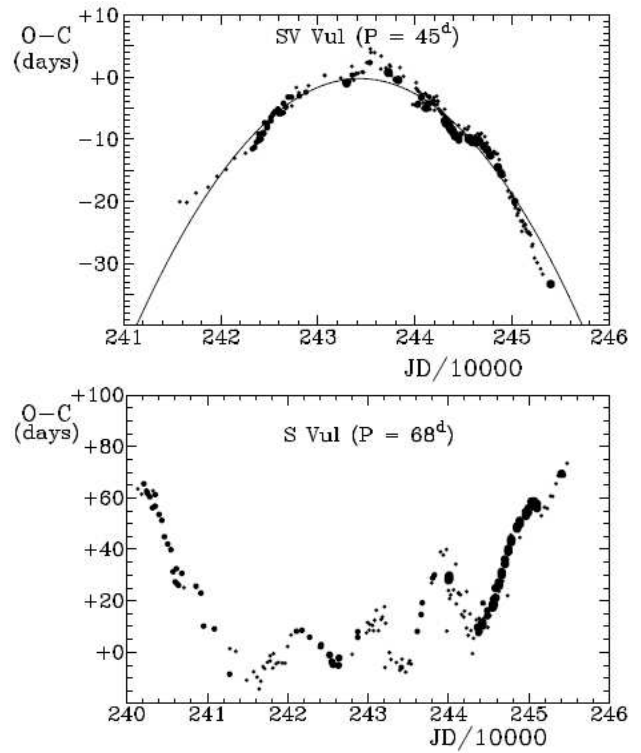}
\caption{Example for secular period decrease (SV~Vul) and 
increase (S~Vul) \citep{Tetal09}}
\label{fig:perchange}
\end{figure}

Intermediate-mass stars cross the classical instability region up to 
5 times during their evolution and the subsequent crossings occur at
increasing luminosities. Due to the slope of the lines of constant 
period in the Hertzsprung-Russell diagram, odd numbered crossings 
result in a secular period increase, while during even numbered 
crossings the pulsation period of Cepheids is decreasing on a long time 
scale (Fig.~\ref{fig:perchange}). The 2nd crossing is the slowest one, 
thus majority of Cepheids are in this evolutionary phase. The observed
rates of period change in over 200 Milky Way  Cepheids were studied
by \citet{Tetal06} for assigning the crossing number. 

Post-main sequence evolution of stars is accompanied with changes in
the surface chemical composition, so in principle, the abundance of
elements derived from the observed spectra supplemented with 
information on the direction of the secular period change is
sufficient for assigning the crossing mode unambiguously. Viability
of such procedure has been presented by the study of SV~Vul
\citep{Tetal04}. Applicability of this method is, however, encumbered
by the fact that the CNO-processed material can be dredged up into
the stellar atmosphere before the red supergiant phase as a consequence
of meridional mixing in rapidly rotating B stars, progenitors of
Cepheids \citep{Tetal06}.

\subsection{Temperature range}
\label{ss:tem} 
  
In fact, the $P$-$L$ relationship is the two-dimensional
projection of the underlying period-luminosity-colour 
($P$-$L$-$C$) relationship. A part of the intrinsic scatter
of the $P$-$L$ relationship is caused by the neglect of the colour 
(temperature) dependence.

Knowledge of $T_{\rm eff}$ is essential for removing 
the reddening effect and correcting for the interstellar 
extinction.

\subsection{Reddening}
\label{ss:red}  
    
Extensive lists of spectroscopically determined
$T_{\rm eff}$ values (and the $E(B-V)$ colour excess derived)
have been compiled by \cite{Aetal05} and quite recently by
\cite{Letal11} and \cite{LL11} involving about 400 Galactic 
Cepheids. For individual calibrating Cepheids in our Galaxy,
the intrinsic colour and apparent brightness can be determined 
from this data base.

The effect of line-of-sight extinction can be practically 
`removed' by using the reddening-free Wesenheit function
for any pair of bands $ij$:
 $$W_{ij}=m_{i}-R_{ij}\times (m_i-m_j)~;$$
\noindent where $R_{ij}=R_i/(R_i-R_j)$. The dereddened $W$ function 
is unaffected by extinction in case the ratio of total-to-selective 
absorption, $R_{ij}$, is universal and known. Such Wesenheit function 
can be defined for any combination of optical and near-IR bandpasses.

This formalism can account for the differential reddening
as long as $R_{ij}$ does not vary from star to star but it cannot
account for the varying amount of dust along the line of sight 
to each Cepheid.

\subsection{Binarity}
\label{ss:bin}  
   
Companion stars effectively contribute to the observed width
of the Cepheid $P$-$L$ relationship. Unresolved companions (either
physical or optical) falsify the brightness and colours of the 
Cepheid to an extent that depends on the luminosity and temperature
differences between the Cepheid and its companion(s). Most of the
bright outliers in Fig.~\ref{fig:empplr} can be unrecognized
binaries in the Large Magellanic Cloud.

As to our Galaxy, an in-depth survey and analysis of the available
observational data indicates that at least 50 per cent of Cepheids 
are not single stars \citep{Sz03}. 

Considering such high frequency of occurrence of binaries 
among Cepheids, it is essential to remove photometric and
other effects of companions when using individual Cepheids for
calibrating the $P$-$L$ relationship.

\begin{figure}[!h]
\includegraphics[width=\columnwidth]{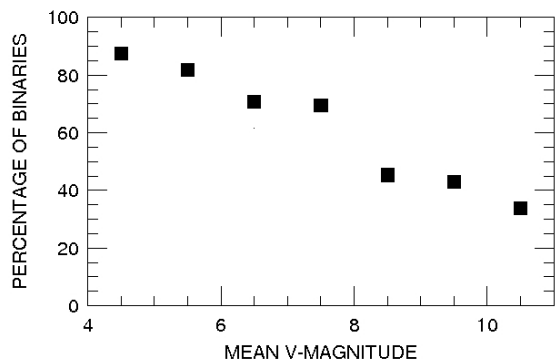}
\caption{Observed frequency of occurrence of binaries among
Galactic Cepheids as a function of the mean apparent
brightness \citep{Szetal11}. The trend seen in the diagram
is the result of observational selection effects}
\label{fig:bincepfreq}
\end{figure}

In fact, it is an observational challenge to reveal binarity of 
fainter Cepheids (cf. the selection effect seen in 
Fig.~\ref{fig:bincepfreq}). There are, however, various methods 
for pointing out faint companions.

\begin{figure}[!ht]
\includegraphics[width=\columnwidth]{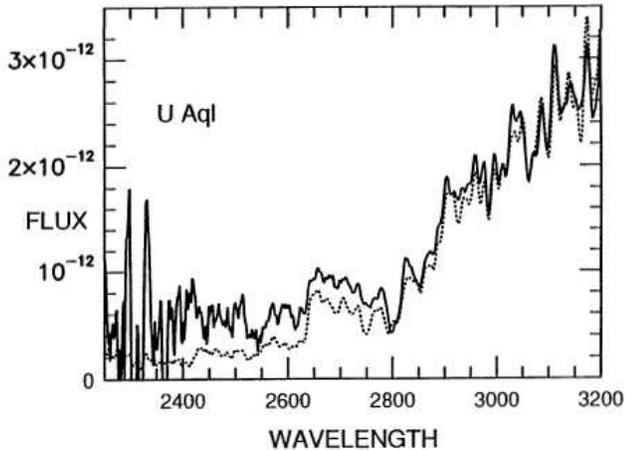}
\caption{Blue companions can be readily detected 
from ultraviolet spectra. U~Aql has a B9.8V type 
secondary \citep{E92}. The dotted curve is the
flux from the Cepheid component}
\label{fig:iuesp}
\end{figure}

{\em Spectroscopy}, especially in the UV region, is
instrumental in revealing faint blue secondaries (see 
Fig.~\ref{fig:iuesp}). Radial velocity phase curves obtained
in different seasons can be used for discovering spectroscopic
binary nature of a Cepheid by detecting the orbital effect
superimposed on the radial velocity variations due to pulsation.

Multicolour {\em photometric} data can also indicate presence of
a companion. \citet{KSz09} devised different parameters based on
the wavelength dependence of the pulsational amplitude which
hint at the presence of a photometric (i.e., not necessarily 
physical) companion.

Companions to Cepheids can be revealed by {\em astrometric} 
methods, as well. The Gaia astrometric space probe (to be
launched in 2013) will be able to resolve dozens of visual binaries
among Cepheids. For some of these systems, even orbital elements
can be determined from the Gaia astrometric measurements spanning
a 5-year-long interval. (The shortest orbital period of a binary
system involving a supergiant Cepheid component is about a year.)

\begin{figure}[!htb]
\includegraphics[width=\columnwidth]{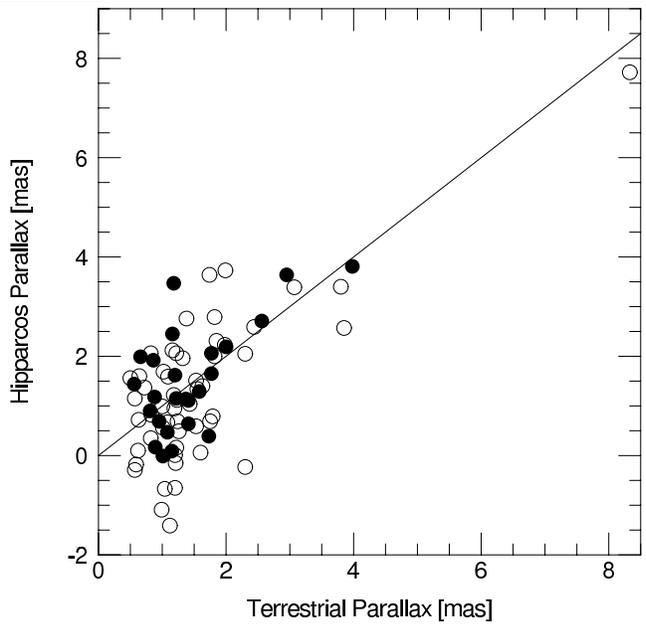}
\caption{Revised Hipparcos parallax of the nearest Cepheids
as a function of their astrophysically determined distances
converted into parallax (`terrestrial parallax'). Filled
circles correspond to Cepheids without known companions, 
empty circles represent Cepheids with companions. It is
remarkable that all negative Hipparcos parallaxes correspond
to Cepheids belonging to binary (multiple) systems \citep{Szetal11}}
\label{fig:parpar}
\end{figure}

The angular sensitivity of about 0.001 arc second of the previous
astrometric satellite, Hipparcos, was not sufficient to resolve
the orbits of binaries containing a Cepheid component. The adverse
effect of the unrecognised orbital motion is, however, obvious even
in the revised Hipparcos parallaxes (Fig.~\ref{fig:parpar}): all
negative parallax values for Cepheids within 2~kpc belong to
Cepheids with known companions, i.e., neglect of the subtle orbital
motion falsified the derived parallax in each case (and in either
sense). If the orbital period exceeds 6-8 years, the astrometric
contribution of the orbital motion is negligible during the 3-5 year 
measurement period of astrometric space missions like Hipparcos or 
Gaia. Some Cepheids marked as single stars in Fig.~\ref{fig:parpar} 
may also have unresolved physical companions as implied by 
Fig.~\ref{fig:bincepfreq}.

If the binary system cannot be resolved even by Gaia astrometry, 
its binarity can be revealed from data to be obtained by the Gaia 
radial velocity spectrometer (to the limiting integral brightness
of about 13th magnitude).

\subsection{Blending}
\label{ss:ble} 

Angular proximity of unrelated stars mainly occurs in crowded
stellar fields, typical of locations of extragalactic Cepheids.
Blending of light from a Cepheid with that of a star along the
same line-of-sight results in a bright outlier in the $P$-$L$ 
relationship. If the disturbing point source cannot be separated 
from the Cepheid, the bright outlier can be considered to be a
binary source involving a Cepheid component (see Sect.~\ref{ss:bin}).
\citet{NK10} presents some examples for both cases among Cepheids
in the Small Magellanic Cloud.

\citet{Petal04} list three methods of handling the problem 
of field-star crowding. In extreme cases, however, the photometric 
contamination cannot be corrected for, and the given Cepheid 
has to be removed from the sample of calibrating stars.

Because long-period Cepheids are more luminous, their photometry is 
less affected by the blending effect.
     
\subsection{Metallicity}
\label{ss:met}  
  
The metallicity sensitivity of stellar luminosity is an obvious
and long-discussed cause of the dispersion of the $P$-$L$ relationship.
The main problem here is that contradictory results have been
achieved from both theoretical calculations and observational data.
For an excellent summary, see \citet{Retal08}.
     
\begin{figure}[!thb]
\includegraphics[width=\columnwidth]{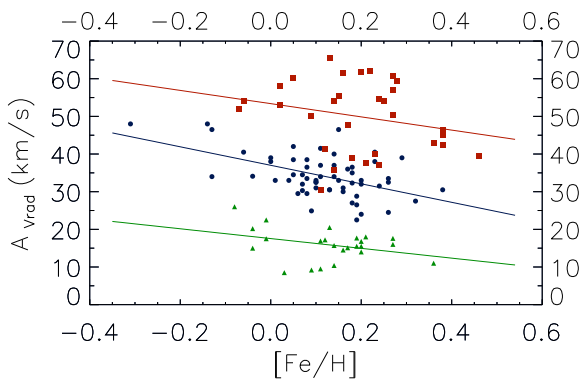}
\caption{Metallicity has an effect on most physical
properties of Cepheids: the peak-to-peak amplitude of
radial velocity variations during the pulsational cycle
also depends on the iron abundance, [Fe/H]. Data are
plotted only for single Cepheids: triangles represent
s-Cepheids, circles short-period Cepheids of normal amplitude, 
squares long-period Cepheids \citep{SzKl11}. Short- and 
long-period Cepheids are separated at the pulsation 
period of 10\fd47}
\label{fig:met}
\end{figure}

Although the debate on the influence of chemical composition on the
$P$-$L$ relationship is unsettled yet, it is undoubted that 
practically all pulsational properties (amplitudes, amplitude ratios) 
of Cepheids depend on the atmospheric metal content \citep{SzKl11}.
It is remarkable that some of these dependences cannot be explained 
by the differential line blanketing. Such an effect is seen in
Fig.~\ref{fig:met}: even the peak-to-peak radial velocity amplitude
slightly depends on the iron content, [Fe/H], of the Cepheid.
An additional interesting fact is that the limiting period that
separates short- and long-period Cepheids appears at 10.47 days, 
instead of the conventional value of 10 days (see Sect.~\ref{ss:nol}).

\subsection{Helium content}
\label{ss:hec}  
  
In addition to metallicity, the helium abundance also makes
influence on the luminosity of the star. Unlike metallicity,
the He content cannot be determined from spectral observations,
so its effect cannot be studied observationally. Theoretical
computations, however, indicate that varying helium/metal
abundance ratio results in different synthetic $P$-$L$ 
relationships, and the effect of compositional differences
is larger for longer pulsation periods \citep{Metal05}.

\subsection{Non-linearity}
\label{ss:nol} 
  
There is growing evidence of the existence of a break in 
the ridge-line $P$-$L$ relationship. The dichotomy between 
short- and long-period Cepheids was pointed out first in the 
LMC. \citet{Setal04} found different slopes of the $P$-$L$ 
relationships of Cepheids with period shorter and longer
than 10 days. Deviations from linearity of the LMC
Cepheid $P$-$L$ and $P$-$L$-$C$ relationships was confirmed
by \citet{Ketal07}.

\begin{figure}[!thb]
\includegraphics[width=\columnwidth]{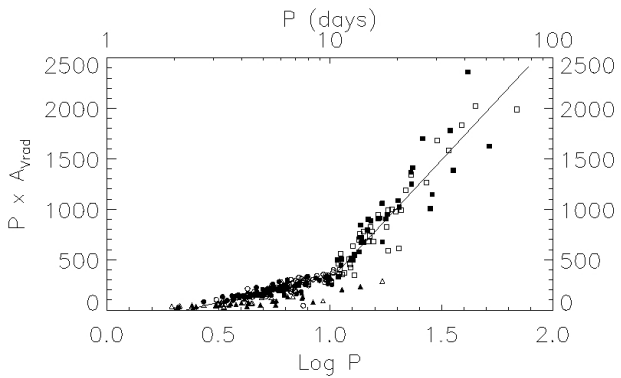}
\caption{This diagram shows a sharp break separating
short- and long-period Cepheids. The dichotomy appears
at $log P = 1.02$. The quantity along the vertical axis,
product of the pulsation period and peak-to-peak radial
velocity amplitude, is a measure of the radius variation
during a pulsational cycle \citep{KSz09}}
\label{fig:break}
\end{figure}

When studying the period dependence of pulsational properties 
of Galactic Cepheids, a similar dichotomy was found by \citet{KSz09}. 
However, as is seen in Fig.~\ref{fig:break},
the break does not appear at an exact value of $P=10$~days, 
in fact, the dichotomy occurs at 10.47 days ($\log P = 1.02$).

A break was revealed in the $P$-$L$ relationship of the
SMC Cepheids, as well \citep{NK10} but, strangely enough,
the dividing period is at $\log P=0.4$.

The presence of such break and the period value separating
short- and long-period Cepheids in the three galaxies (Milky Way, 
LMC, SMC) implies that the dichotomy is not a metallicity effect
but the location of the break can be metallicity dependent.

\subsection{Miscellaneous other effects}
\label{ss:other} 

There are other physical effects, e.g., stellar magnetic field, 
overshooting, etc. that also contribute to the scatter in the 
$P$-$L$ plot constructed from observational data. 

In the case of the nearest galaxies the line-of-sight extent
of the system is an additional factor leading to increased
scatter of the points in the $P$-$L$ diagram. This effect can be 
corrected for, as was successfully done by \citet{Netal04} and
\citet{Petal04} in constructing the $P$-$L$ plot for Cepheids 
in the Large Magellanic Cloud. This geometric effect of 
back-to-front distance differences of Cepheids in remote galaxies, 
especially outside the Local Group can be neglected.

\section{Conclusion}
\label{s:concl}

More than ten factors contributing to the scatter in the 
$P$-$L$ relationship have been listed and removal of some
of these effects has been discussed.
All these effects are independent of each other. Therefore,
neglect of some factors falsify results on the role of other 
factors in observational studies. For example, when studying 
the effect of metallicity on stellar luminosity, Cepheids 
belonging to binary systems have to be excluded.
Another example is the determination of the tilt of the
Large Magellanic Cloud from the dispersion of the $P$-$L$
relationship. In this procedure, the deviation of individual
Cepheids from the ridge-line $P$-$L$ relationship is usually
attributed solely to the depth effect, neglecting effects of
metallicity and binarity, etc. 

Reduction of the scatter can be achieved by turning to longer
wavelengths. However, moving to infrared spectral region does 
not remove the whole spread in the $P$-$L$ relationship.
The effect of the interstellar extinction becomes negligible 
in near-IR but effect of metallicity increases (and reverses)
toward longer infrared wavelengths \citep{FM11}.
Moreover, circumstellar envelopes that commonly occur around 
Cepheids cause an IR excess \citep{Betal11}.

In view of the increasing number of extragalactic Cepheids, the
non-linearity of the $P$-$L$ relationship can be and has to be 
studied beyond the Magellanic Clouds. The behaviour of the 
long-period Cepheids is especially important because such stars 
overwhelm among Cepheids known in remote galaxies.

\acknowledgments

This research has been supported by the ESA PECS 98090 project.
Constructive remarks by the referee are gratefully acknowledged.


\begin{thebibliography}{}

\bibitem[\protect\citeauthoryear{Andrievsky et~al.}{2005}]{Aetal05}
Andrievsky S.~M., Luck R.~E., \& Kovtyukh V.~V. 2005, \aj, 130, 1880
\bibitem[\protect\citeauthoryear{Barmby et~al.}{2011}]{Betal11}
Barmby P., Marengo M., Evans N.~R., Bono G., Huelsman D., Su K.~Y.~L., 
Welch D.~L., \& Fazio G.~G. 2011, \linebreak \aj, 141:42 
\bibitem[\protect\citeauthoryear{Evans}{1992}]{E92}Evans N.~R. 1992, 
\apj, 384, 220
\bibitem[\protect\citeauthoryear{Freedman \& Madore}{2010}]{FM10}
Freedman W.~L. \& Madore B.~F. 2010, \araa, 48, 673
\bibitem[\protect\citeauthoryear{Freedman \& Madore}{2011}]{FM11}
Freedman W.~L. \& Madore B.~F. 2011, \apj, 734, 46
\bibitem[\protect\citeauthoryear{Klagyivik \& Sza\-bados}{2009}]{KSz09}
Klagyivik P. \& Szabados L. 2009, \aap, 504, 959
\bibitem[\protect\citeauthoryear{Koen et~al.}{2007}]{Ketal07}
Koen C., Kanbur S., \& Ngeow C. 2007, \mnras, 380, 1440
\bibitem[\protect\citeauthoryear{Luck \& Lambert}{2011}]{LL11}
Luck R. E. \& Lambert D.~L. 2011, \aj, 142, 136
\bibitem[\protect\citeauthoryear{Luck et~al.}{2011}]{Letal11}
Luck R. E., Andrievsky S. M., Kovtyukh V.~V., Gieren W.~P., 
\& Graczyk D. 2011, \aj, 142:51
\bibitem[\protect\citeauthoryear{Marconi}{2009}]{M09}
Marconi M. 2009, \memsai, 80, 141
\bibitem[\protect\citeauthoryear{Marconi et~al.}{2005}]{Metal05}Marconi M.,
Musella I., \& Fiorentino G. 2005, \apj, 632, 590
\bibitem[\protect\citeauthoryear{Ngeow \& Kanbur}{2010}]{NK10}
Ngeow C.-C. \& Kanbur S.~M. 2010, \apj, 720, 626
\bibitem[\protect\citeauthoryear{Nikolaev et~al.}{2004}]{Netal04}
Nikolaev S., Drake A.~J., Keller S.~C., Cook K.~H., Dalal~N.,
Griest K., Welch D.~L., \& Kanbur S.~M. 2004, \linebreak \apj, 601, 260
\bibitem[\protect\citeauthoryear{Og{\l}oza et~al.}{2000}]{Oetal00}
Og{\l}oza W., Moskalik P., \& Kanbur S. 2000, in Proc. IAU Coll. 176, The 
Impact of Large-Scale Surveys on Pulsating Star Research, ASP Conf. Ser. 
203, Ed. L. Szabados \& D.~Kurtz (San Francisco: ASP), p.\,235
\bibitem[\protect\citeauthoryear{Persson et~al.}{2004}]{Petal04}
Persson S.~E., Madore B.~F., Krzemi\'nski~W., Freedman W.~L., Roth~M.,
\& Murphy D.~C. 2004, \aj, 128, 2239
\bibitem[\protect\citeauthoryear{Romaniello et~al.}{2008}]{Retal08}
Romaniello M., Primas F., Mottini M., Pedicelli S., Lemasle B., Bono~G., 
Fran\c{c}ois~P., Groenewegen~M.~A.~T., \& Laney~C.~D. 2008, \aap, 488, 731
\bibitem[\protect\citeauthoryear{Sandage et~al.}{2004}]{Setal04}
Sandage A., Tammann G.~A., \& Reindl B. 2004, \aap,  424, 43
\bibitem[\protect\citeauthoryear{Soszy\'nski et~al.}{2008}]{Setal08}
Soszy\'nski I., Poleski R., Udalski A., Szyma\'nski M.~K.,
\linebreak Kubiak M., Pietrzy\'nski G., Wyrzykowski {\L}., Szewczyk O., 
\& Ulaczyk K. 2008, \actaa, 58, 163
\bibitem[\protect\citeauthoryear{Soszy\'nski et~al.}{2010}]{Setal10}
Soszy\'nski I., Poleski R., Udalski A., Szyma\'nski M.~K.,
\linebreak Kubiak M., Pietrzy\'nski G., Wyrzykowski {\L}., Szewczyk O.,
 \& Ulaczyk K. 2010, \actaa, 60, 17
\bibitem[\protect\citeauthoryear{Szabados}{2003}]{Sz03}
Szabados L. 2003, Inf. Bull. Var. Stars, No.\,5394
\bibitem[\protect\citeauthoryear{Szabados \& Klagyivik}{2011}]{SzKl11}
Szabados L. \& Klagyivik P. 2011, accepted by \aap, (arXiv:1112.0115)
\bibitem[\protect\citeauthoryear{Szabados et~al.}{2011}]{Szetal11}
Szabados L., Kiss Z. T., \& Klagyivik P. 2011, EAS Pub. Ser., 45, 441
\bibitem[\protect\citeauthoryear{Szab\'o et~al.}{2007}]{Szetal07}
Szab\'o, R., Buchler, J.~R., \& Bartee J. 2007, \apj, 667, 1150
\bibitem[\protect\citeauthoryear{Turner \& Berdnikov}{2004}]{Tetal04}	
Turner D. G. \& Berdnikov L.~N. 2004, \aap, 423, 335
\bibitem[\protect\citeauthoryear{Turner et~al.}{2006}]{Tetal06}	
Turner D. G., Abdel-Sabour Abdel-Latif M. \& Berdnikov L.~N. 2006,
\pasp, 118, 410
\bibitem[\protect\citeauthoryear{Turner et~al.}{2009}]{Tetal09}	
Turner D. G., Majaess D.~J., Lane D.~J., Szabados L., \linebreak 
Kovtyukh V.~V.,
Usenko I.~A. \& Berdnikov L.~N. 2009, in Stellar Pulsation: Challenges for
Theory and Observation, AIP Conf. Proc., 1170, 108
\end{thebibliography}
\end{document}